\documentstyle[aps]{revtex}
\begin{document}

\draft
\title{Spectral Flow  in Vortex Dynamics of
$^3$He-B and Superconductors.
 }
\author{ N.B. Kopnin $^{(1,2)}$,  G.E. Volovik $^{(1,2)}$, and
\"U. Parts $^{(1)}$}

\address{$^{(1)}$  Low Temperature Laboratory,
 Helsinki University of Technology.\\
 Otakaari 3A, 02150 Espoo, Finland.\\
$^{(2)}$  L.D. Landau Institute for Theoretical Physics,
 Kosygin Str. 2, 117940 Moscow, Russia }

\date{\today}
\maketitle

\begin{abstract}
Nondissipative force acting on a vortex in Fermi superfluids and
superconductors is calculated from the spectral flow of fermion zero modes
in the vortex core. The spectral flow effect casts a new light on the
problem of mutual friction in superfluids. We demonstrate that, for
vortices in an isotropic system, the reactive mutual friction parameter
D'(T) is negative in the collisionless regime and it is positive in the
hydrodynamic regime. This agrees with 3He-B measurements [1] ( T.D.C.
Bevan, et al,  Phys. Rev. Lett., 74, 750 (1995)), which thus give the
experimental verification of the spectral flow effect in the vortex
dynamics. General expression for  all 3 nondissipative forces (Magnus,
Iordanskii and spectral flow) is presented in the whole temperature range.
\end{abstract}

%\narrowtext
%\twocolumn
\baselineskip=5.4mm

Spectral flow of fermions is the topological phenomenon related to
anomalies in the relativistic quantum field theory. The inhomogeneous vacuum
induces a spectral flow, which carries particles
from negative energy states of the  vacuum to positive energy states.
As a result, the corresponding conserved quantity (viz, charge) is
transferred from a coherent
vacuum motion into incoherent fermionic degrees of freedom, which is
visualized  as creation of charge from the vacuum. It was shown in
\cite{A-phase anomaly,Stone}
that the same phenomenon takes place in superfluid $^3$He-A, where the gap
in the
quasiparticle spectrum has topologically stable point nodes, resulting in a
spectral flow of Bogoliubov quasiparticles through the nodes when the
condensate evolves in time. This leads to
transfer of a linear momentum from the coherent   superfluid motion into the
normal component of the liquid, i.e., to a force between the
superfluid and normal components.

A mutual friction force of the same type arises
for a Fermi superfluid  or superconductor even without
gap nodes, if quantized vortices are present \cite{Q-modes-Index}.
The singularity at the vortex axis plays the
same role as gap nodes in  A-phase of $^3$He, while the
evolution of the condensate, which leads to the spectral
flow, is realized by the motion of the vortex with respect
to the heat bath. A microscopic analysis of the spectral
flow in vortices was also presented in \cite{Gaitan}, but the
hydrodynamic and collisionless regimes were not distinguished there which led
to the contradictory conclusion.  Here we consider the influence of the
spectral flow on a low-dissipative vortex dynamics in both regimes for
isotropic Fermi systems such as
$s$-wave superconductors and $^3$He-B. For superconductors, we consider a simple
case  of Galilean invariant system, i.e., we ignore energy-band effects in
metals
and  assume that the mean free path is longer
than the coherence length (clean case).
We compare the theoretical mutual friction force
in $^3$He with the experimental results of Ref.\cite{Bevan}. The observed
change
of sign of the parameter $D^\prime$ following  the change of regimes
illustrates the essential contribution of
the spectral flow effect into the vortex dynamics.

{\it Nondissipative forces in Bose superfluids.}
There is a big difference between dynamics of  quantized
vorticity in Bose and Fermi superfluids. In Bose liquids the nondissipative
force acting on the vortex moving with respect to superfluid and normal
components of the liquid is simply a sum of the Magnus forces
from the superfluid and normal components:
\begin{equation}
{\bf F}_{nd }=\kappa\hat {\bf  z}\times{\breve \rho}_s(T)({\bf v}_L-{\bf
v}_s)+\kappa\hat {\bf  z}\times    {\breve \rho}_n(T)({\bf v}_L-{\bf v}_n)
.\label{1.2}
\end{equation}
Here  ${\bf v}_L$ is the vortex velocity,   ${\bf v}_s$ and ${\bf v}_n$
are the superfluid and normal velocities; $\kappa\hat {\bf  z}$ is the
circulation vector. For Bose  superfluids  $\kappa=2\pi N\hbar /m$, where
$N$ is the vortex winding number, and $m$ is the bare mass of a boson. The
densities ${\breve \rho}_s(T)$ and ${\breve \rho}_n(T)$ of the normal and
superfluid components, respectively, can be  anisotropic tensors.

Eq. (\ref{1.2}) can be rearranged to comply with the  Landau picture of a
superfluid, where the motion  of a
superfluid vacuum of the total density $\rho$ with the superfluid velocity
${\bf v}_s$  is supplemented with the dynamics of  elementary excitations
moving with the velocity ${\bf v}_n-{\bf v}_s$. The force becomes
\begin{eqnarray}
{\bf F}_{nd}
=\kappa\hat {\bf  z}\times \rho({\bf v}_L-{\bf v}_s)+\kappa\hat {\bf  z}\times
{\breve
\rho}_n(T)({\bf v}_s-{\bf v}_n) \nonumber\\
={\bf F}_{\rm Magnus}+{\bf F}_{\rm
Iordanskii} \label{1.1}
\end{eqnarray}
Now the   Magnus force is defined as if the vortex moves with respect to the
superfluid vacuum.  The Iordanskii force \cite{Iordanskii,Sonin1} results
from elementary excitations outside the vortex core:   the vortex line
provides them with an Aharonov-Bohm  potential.

The Iordanskii force in Eq. (\ref{1.1}) is
a sum of elementary forces which act on individual particles,
 $\partial_t{\bf p}= ({\bf \nabla}\times{\bf v}_s)\times{\bf p}$. Here ${\bf
p}$ is the quasiparticle momentum and the vorticity ${\bf \nabla}\times{\bf v}_s
=\kappa\hat {\bf  z}\delta_2({\bf r})$ is concentrated in the vortex core
"tube". One has
\begin{eqnarray}
- \sum_{{\bf p}} f \partial_t{\bf p}=-\int { d^3p\over (2\pi^3)}\,
\kappa\hat {\bf  z}\times  {\bf p}\,f[E_{{\bf  p}}+{\bf p}\cdot ({\bf v}_s-{\bf
v}_n)]\nonumber \\ =\kappa \hat {\bf  z}\times {\breve \rho}_n(T)({\bf v}_n-{\bf
v}_s) .\label{1.4}
\end{eqnarray}
Here $f(E)$ is the distribution function of elementary excitations, Doppler
shifted due to the counterflow ${\bf v}_n-{\bf v}_s$. The Iordanskii force
has the
same origin as the Aharonov-Bohm  effect for spinning cosmic strings
\cite{3Forces,CausalityViolation}.

Adding  a dissipative friction force  to Eq.\,(\ref{1.2}), one obtains a
conventional expression for the  balance of forces acting on the vortex
\cite{Donnelly}
\begin{equation}
\kappa\rho_s({\bf v}_s-{\bf v}_L)\times\hat {\bf  z}+D({\bf v}_n-{\bf
v}_L)-D^\prime({\bf v}_n-{\bf v}_L)\times\hat {\bf  z}  =0 .
\label{1.5}
\end{equation}
Here the term with $D$  is  associated with the friction while the term with
$D^\prime$ is due to the reactive force ${\bf F}_{nd}$.  Comparison with
Eq.(\ref{1.2}) gives the reactive parameter
\begin{equation}
D^\prime=-\kappa {\breve \rho}_n(T) .\label{1.6}
\end{equation}
Our consideration is valid   when dissipation can be neglected compared to
the reactive terms, $D\ll D^\prime$. It is only under this condition the
nondissipative forces do
not depend on details of the core structure and of the quasiparticle kinetics,
and are defined by general characteristics of the vortex and
the superfluid.  Eq.(\ref{1.6}) coincides with microscopic calculations
(see review
\cite{SoninReview}).

 {\it Nondissipative forces in Fermi superfluids.}
Now we consider Fermi systems such as superfluid phases of $^3$He
and  superconductors. For the latter we assume
a large magnetic-field penetration depth compared to the intervortex distance:
in this case the magnetic field can be treated as homogeneous.
We also assume that the magnetic induction $B$ is much smaller than the upper
critical field $H_{c2}$, so that vortex cores do not overlap.

In Fermi superfluids, there is a new contribution to ${\bf F}_{nd}$ in addition
to  Eq.\,(\ref{1.1}) (now $\kappa= \pi N\hbar /m$ and $m$ is
the bare mass of a fermion):
\begin{equation}
{\bf F}_{nd}={\bf F}_{\rm Magnus}+{\bf F}_{\rm Iordanskii}+
{\bf F}_{\rm sp.flow} ~~,\label{2.1}
\end{equation}
This contribution  is related to the chiral fermion zero modes, which are absent
in  Bose superfluids.

The spectrum of single-fermionic excitations in a vortex, $E_n(p_z,Q)$,
depends on the momentum projection $p_z$ on the vortex axis;   the orbital
quantum number $Q$, integer or half of odd
integer, which corresponds to the generalized angular momentum conserved in an
axisymmetric vortex; and $n$ denotes radial
and spin quantum numbers. The interlevel distance of bound states
$\partial E_n/\partial Q=\omega _n$ is small compared to the gap
amplitude $\Delta(T)$ of fermions in bulk:
$\omega_n\sim \Delta^2(T)/E_F$. This spectrum has anomalous
(chiral) branches $E_0(p_z,Q)$ whose number $N_{\rm zm}$ is
related to the vortex winding number  $N_{\rm zm}=2N$ according to the index
theorem \cite{Q-modes-Index}.  As a function of $Q$, each anomalous branch
crosses zero of energy an odd number of times and runs through both discrete
and contunuous spectrum from $E_0=-\infty $ to $E_0=+\infty $.
Any other branch either
does not cross zero of energy at all or crosses it  an even number of times. For
low-energy bound states, the spectrum of the chiral branch is linear in $Q$.
For the most symmetric  vortices, for example, \cite{Caroli}
\begin{equation}
E_0(p_z,Q)=Q\omega_0(p_z) .\label{2.2}
\end{equation}
Due to an odd number of crossings of zero, the chiral branch can result in a
momentum exchange between
fermions localized in the vortex core and fermions in the heat bath. The
corresponding contribution to the force
thus depends on ${\bf v}_n-{\bf v}_L$. In contrast to the two other forces, this
contribution does depend on the core structure and on the quasiparticle
kinetics:
It is determined by the parameter
$\omega_0\tau$, where  $\tau$ is the lifetime of fermions. Two limiting cases
are important, the
collisionless  and hydrodynamic limits: they correspond to $\omega_0\tau\gg 1$
and $\omega_0\tau\ll 1$, respectively. In both  cases dissipation is
small \cite{Kopnin-Kravtsov1,Kopnin,Kopnin-Lopatin} and the results
should thus not be sensitive to the core structure of the vortex.

 {\it Hydrodynamic limit.}
In the hydrodynamic limit $\omega_0\tau\ll 1$ the interlevel spacing is smaller
than the level width $1/\tau$ and the spectrum  $E_0(p_z,Q)$ can be
considered as a function of the continuous parameter $Q$; it crosses zero
at some value of $Q$ [at $Q=0$ for the spectrum of Eq. (\ref{2.2})].
When the vortex moves with respect to the
heat bath (normal component), the velocity difference ${\bf v}_n-{\bf v}_L$
induces a flow of quasiparticles from negative levels to  positive levels
of the spectrum  $E_0(p_z,Q)$. Since the angular momentum
evolves as $Q\rightarrow Q+({\bf r}(t)\times {\bf
p})\cdot{\hat {\bf z}}=Q+ t(({\bf v}_L-{\bf v}_n)\times {\bf
p})\cdot{\hat {\bf z}}$, the number of levels crossing zero energy per unit time
is
$\partial_tQ=({\bf v}_L-{\bf v}_n)\cdot ({\bf p}\times{\hat {\bf z}})$. Each
level bears the linear momentum ${\bf p}$, therefore, the total flux of the
linear
momentum from the vortex to the heat bath is
\begin{eqnarray}
\partial_t{\bf P}=\sum {\bf p}~(-{\partial f\over \partial Q})\partial_t Q
= -{1\over 2}\sum_{n,Q} {{\partial f(E_n)}\over {\partial Q}}\nonumber \\
\times \int_{-p_F}^{p_F} {{dp_z}\over{2\pi}}
\int_0^{2\pi}{{d\phi}\over{2\pi}}~{\bf p}~[ (({\bf v}_L-{\bf v}_n)\times {\bf
p})\cdot{\hat {\bf z}}]\nonumber \\ =\pi N {{p_F^3}\over
{3\pi^2}}{\hat {\bf z}}\times({\bf v}_L-{\bf v}_n).\label{2.3}
\end{eqnarray}
Here we used that only zero modes contribute the sum $
-\sum_{n,Q} (\partial f(E_n)/\partial Q)=\sum_n
(f(E_n=-\infty)-f(E_n=\infty))=2N$  as $Q$ together with
$E_0(Q) $ run from $-\infty$ to
$+\infty $.  Thus the spectral flow contribution to
${\bf F}_{nd}$  in  the hydrodynamic limit is
\begin{equation}
{\bf F}_{\rm sp.flow}=\kappa\hat {\bf  z}\times C_0({\bf v}_n-{\bf v}_L)
,~\omega_0\tau\ll 1,\label{2.4}
\end{equation}
where the anomaly parameter $C_0$ is expressed in terms of the  Fermi momentum
$p_F$ and coincides with the mass density of the normal state:
\begin{equation}
C_0=mp_F^3/3\pi^2~~.\label{2.5}
\end{equation}
The  spectral flow of the fermion zero modes within the vortex core   is
a realization of the Callan-Harvey mechanism of the anomaly cancellation
in the relativistic quantum field theories \cite{Q-modes-Index,Callan}.

 {\it Collisionless limit in uncharged Fermi superfluids}.
In the collisionless limit $\omega_0\tau\gg 1$ the level width is much smaller
than   the interlevel distance and the spectral flow along the discrete levels
is suppressed. The spectral flow becomes possible only for  $E_0(p_z,Q)$
above the gap in the region of the continuous spectrum.
Since only the levels in the range
$\Delta(T)<E_0(p_z,Q)< \infty $ are "fluid" the spectral flow takes place
only at
nonzero $T$ due to the thermal tail of Fermi function $f(E_n)$ in
Eq.(\ref{2.3}).
It is reduced by the factor $[f(-\Delta(T))-f(-\infty)]-[f(\Delta(T))-f(\infty)]
=1-\tanh [\Delta(T)/2T]$
as compared to the hydrodynamic limit (here we assume that the gap is
isotropic).

Thus for an isotropic pair-correlated system, such as $^3$He-B, which has an
isotropic gap $\Delta(T)$,
\begin{equation}
{\bf F}_{\rm sp.flow}=\kappa C_0\, \Bigl[1-\tanh{{\Delta (T)}\over {2T}}\Bigr]~
\hat {\bf  z}\times ({\bf v}_n-{\bf v}_L),
{}~\omega_0\tau\gg 1 .\label{2.6}
\end{equation}

 {\it Spectral flow in superconductors.}
For a charged Fermi-system the levels above the gap are also discrete
due to  quantization in the magnetic field, the interlevel distance being
the cyclotron frequency $\omega_c$. Note that $\omega_c\ll \omega _0$
due to the condition $B\ll H_{c2}$. The levels can be considered as a continuum
only under the condition $\omega_c\tau\ll 1$. The results Eqs.(\ref{2.4}) and
(\ref{2.6}) are thus reproduced for superconductors in the limit
$\omega_c\tau\ll 1$. In the extreme collisionless limit, when
$\omega_c\tau\gg 1$, the spectrum is discrete everywhere and the spectral
flow is completely suppressed. Hence,
\begin{eqnarray}
{\bf F}_{\rm sp.flow}= \kappa C_0\hat {\bf  z}\times ({\bf v}_n-{\bf v}_L),
{}~\omega_0 \tau \ll 1,~\omega_c \tau \ll 1;\label{2.7}\\
{\bf F}_{\rm sp.flow}=\kappa C_0~(1-\tanh{\Delta(T))\over
2T})~\hat {\bf  z}\times({\bf v}_n-{\bf v}_L) ,\nonumber \\
\omega_0\tau\gg 1,~\omega_c\tau\ll 1;\label{2.8}\\
{\bf F}_{\rm sp.flow}=0,~\omega_0\tau\gg 1,~\omega_c\tau\gg 1.
\label{2.9}
\end{eqnarray}

 {\it  Parameter $D^\prime $ for $^3$He-B}.
Let us  consider first the effect of the spectral flow on the
mutual friction parameter $D'$ in $^3$He-B.
The dissipative force, proportional to $D$, is small in both
limits, $\omega_0\tau\ll 1$ and $\omega_0\tau\gg 1$\cite{Kopnin}.
Comparing Eq.(\ref{1.5}) with Eqs.(\ref{1.1},\ref{2.1},\ref{2.4}),
we find that the reactive parameter in the hydrodynamic limit is
\begin{equation}
D^\prime_{\rm hydro}=\kappa [C_0-\rho_n(T)] ,~\omega_0\tau\ll
1.\label{3.1}
\end{equation}
Here it is used that the normal density $\rho_n$ in $^3$He-B is isotropic.
In the
collisionless limit we find from Eq. (\ref{2.6})
\begin{equation}
D^\prime_{\rm coll}=\kappa \Bigl[ C_0\Bigl(1-
\tanh{\Delta(T)\over 2T}\Bigr)-\rho_n(T)\Bigr], ~\omega_0\tau\gg 1.\label{3.2}
\end{equation}
Eq.(\ref{3.1}) was obtained in \cite{Q-modes-Index} in the limit $T\rightarrow
0$ and was extended to nonzero $T$ in \cite{3Forces,Makhlin-Misirpashaev}.
In the weak coupling approximation the anomaly parameter $C_0\approx \rho$,
since $(\rho -C_0)\approx \rho \Delta^2(T)/E_F^2\ll \rho$.

If one neglects the difference between $C_0$ and $\rho$,
\begin{eqnarray}
D^\prime_{\rm hydro} \kappa \rho_s(T) ,~\omega_0\tau\ll 1,\label{3.4}\\
D^\prime_{\rm coll} \kappa \Bigl[\rho_s(T)-\rho  \tanh{\Delta(T)\over 2T}\Bigr] ,
{}~\omega_0\tau\gg 1 .\label{3.5}
\end{eqnarray}
Eq. (\ref{3.4}) in the limit $T\ll T_c$ was derived for superconductors in
\cite{Kopnin-Kravtsov1} and for $^3$He in \cite{Kopnin}; it was extended for
higher temperatures in \cite{Kopnin-Lopatin}.
Eq. (\ref{3.5}) was first obtained for superconductors in
\cite{Kopnin-Kravtsov2,GalpSonin} by calculating the force produced by
excitations scattered by moving vortex. The same result has been reproduced
in \cite{Kopnin-Lopatin} using the kinetic equation.
The classical results of Bardeen and Stephen\cite{Bardeen} and of
Nozieres and Vinen\cite{Nozieres} can also be compared with Eqs. (\ref{3.4})
and (\ref{3.5}):
The former corresponds to the  hydrodynamic limit $D^\prime =\kappa\rho_s$. The
result   of \cite{Nozieres}, namely that the only force left  at
$T=0$ is the Magnus force, remains valid in a collisionless limit, when both
the Iordanskii force and the spectral flow disappear. Its generalization which
extrapolates  between the two limits
\cite{Vinen-Warren} does not contain the spectral flow term
$1-\tanh [\Delta(T)/ 2T]$ and thus is valid only at
$T=0$.

Applying the Eqs.(\ref{3.4},\ref{3.5}) to $^3$He-B one finds that these two
regimes take place in the regions $T\ll T_c$ and  $T_c-T\ll T_c$. For $T$ close
to $T_c$,  the hydrodynamic limit is always realized since
$\omega_0\sim\Delta^2(T)/E_F\rightarrow 0$, and the reactive parameter is
positive $D^\prime\approx \kappa\rho_s(T)$. At  low $T\ll \Delta(0)$ the
collisionless limit is always reached, hence the parameter
$D^\prime $ is negative according to Eq.(\ref{3.5}).  Indeed, the
ratio of $D^\prime $ values for these two regimes is
\begin{equation}
D^\prime_{\rm coll}/D^\prime_{\rm hydro}
=1-{\rho\over\rho_s(T)}\tanh{\Delta(T)\over 2T}~<~0~.\label{3.6}
\end{equation}
This dependence is shown in Fig.1. The sign reversal of
$D^\prime$ is clearly seen in the oscillating membrane experiment
in $^3$He-B \cite{Bevan}.

Using the results of Ref. \cite{Kopnin-Lopatin}, we can write a phenomenological
expression which interpolates between  the hydrodynamic and collisionless
limits:
\begin{eqnarray}
D\approx \kappa\rho_s {\omega_0\tau  \over 1+\omega_0^2\tau^2};\\
D^\prime\approx \kappa \Bigl[ C_0-\rho_n(T)-{\omega_0^2\tau^2  \over 1+
\omega_0^2\tau^2}~C_0~\tanh{\Delta(T)\over 2T}~\Bigr] .\label{3.7}
\end{eqnarray}
The  temperature dependences of $D(T)$ and $D^\prime(T)$ can be visualized
using a simple model expression for the lifetime \cite{Kopnin} $\tau \sim
(E_F/T^2)\exp [\Delta (T)/T]$,  so that the parameter $\omega _0\tau $
is approximated by
\begin{equation}
\omega _0\tau  = \alpha [\Delta (T)/T]^2\exp [\Delta (T)/T], \label{3.8}
\end{equation}
where $\alpha$ is a fitting constant.
The plot in  Fig.~2  qualitatively agrees with the experimental temperature
dependence found in
\cite{Bevan}. We see that  $D^\prime/ \kappa  \rho_s$ approaches its
negative collisionless-limit asymptote of Eq.(\ref{3.6}) at low $T$ and
tends to its positive
hydrodynamic-limit asymptote $D^\prime/ \kappa  \rho_s =1$ at high $T$. The
observed sign reversal of $D^\prime$ is the experimental illustration of the
spectral flow  force in the vortex dynamics.
Eq. (\ref{3.7}) suggests that for $T\rightarrow 0$, one always has
$D^\prime \approx -\kappa \rho _n$ since $\rho _n/\rho \gg
1-\tanh [\Delta (0)/2T]$, and $(\omega _0\tau )^{-2}$ decreases faster than
$\rho _n$.

{\it Superconductors}.
Eq. (\ref{3.7})  can  also be used for investigation of
the sign of the Hall effect in superconductors. The Hall conductivity
in the limit $\omega_c\tau \ll 1$ is
\begin{eqnarray}
\sigma_H= c^2 (\kappa\rho_s- D^\prime)/\Phi _0B\nonumber \\
\approx
{{ec}\over {mB}}
\Bigl[ \rho -C_0+ C_0{\omega_0^2\tau^2  \over 1+\omega_0^2\tau^2}
\tanh{\Delta(T)\over2T}\Bigr]  .\label{3.9}
\end{eqnarray}
where $\Phi _0=\pi c/e$ is the magnetic flux quantum.
If $C_0>\rho$ the sign  reversal of $\sigma_H$ can occur in the
hydrodynamic regime as proposed in Ref.\cite{Feigel'man} .

 {\it Conclusion.}
We found a spectral flow contribution to the nondissipative force acting
on a moving vortex in hydrodynamic and collisionless limits. Due
to the spectral flow the reactive mutual friction parameter $D^\prime$ has
different signs in these two limits. This sign reversal was observed in the
$^3$He-B experiments on the transition between the two regimes,
realized at different temperatures. This confirms the
essential role of the spectral flow effect in the vortex dynamics of Fermi
superfluids and supercoductors.

We thank Yu. G. Makhlin and T.Sh. Misirpashaev
for   useful discussions. This work was supported by the  Academy of Finland
and by the Russian Foundation for Fundamental Research, Grants No. 93-02-02687
and 94-02-03121.

\begin{figure}
\caption{ Theoretical dependence of the ratio
$ D^\prime_{\rm coll}/ D^\prime_{\rm hydro}$
from Eq. (\protect\ref{3.6}) for an isotropic superfluid.
This ratio is always negative. The gap function $\Delta(T)$ and the superfluid
density $\rho_s(T)$ are taken in the weak-coupling BCS approximation.
$F_1$ is the Fermi-liquid parameter which enters $\rho _s$
in a Fermi superfluid. For $^3$He, the values of $F_1$ correspond to pressures
$p=0$, $p=12$, and $p=24$ bar, respectively.
 }  \end{figure}

\begin{figure}
\caption{Qualitative temperature dependences of $D/\kappa \rho _s$
and $D^\prime /\kappa \rho _s$ from Eq. (\protect\ref{3.7}) in a simple
model of Eq.(\protect\ref{3.8}). The fitting parameter $\alpha =0.02$.
 }
\end{figure}

\end{document}